# On Delayed Choice and Contingent Absorber Experiments

R. E. Kastner

rkastner@umd.edu

University of Maryland, College Park

ABSTRACT. It is pointed out that a slight variation on the Wheeler Delayed Choice Experiment presents the same challenge to orthodox quantum mechanics as Maudlin-type contingent absorber experiments present to the Transactional Interpretation (TI). Therefore, the latter cannot be used as a basis for refutation of TI.

1. Introduction and Background

This paper discusses the famous 'Delayed Choice Experiment" (DCE) proposed by Wheeler (1978), along with 'Contingent Absorber Experiments' (CAE) proposed as challenges for the Transactional Interpretation of quantum mechanics (TI) of Cramer (1986). While the DCE itself is well known, the implications of the DCE for possible ontologies underlying quantum theory have not been fully explored. Some of these implications are particularly relevant to a thought experiment first proposed by Maudlin (1996) that has been generally taken as a refutation of TI.

TI is a time-symmetric interpretation in which outcomes (and their associated probabilities) supervene on an interaction between an emitter and all absorbers accessible to it. In TI, the emitter emits an "offer wave" (OW) corresponding to the standard (retarded) solution to the Schrodinger Equation, and the absorbers respond to that OW by generating advanced "confirmation waves" (CW) which correspond to solutions of the complex conjugate Schrodinger equation. The Maudlin experiment can be termed a 'Contingent Absorber Experiment' (CAE) in that the placement of at least one absorber is contingent on a previous outcome between the emitter and a nearby absorber. The contingent nature of the more distant absorber placement (together with its various implications) has been seen as a serious challenge to the consistency of the TI approach. It is argued here that the DCE, with a slight variation, presents essentially the same challenge to standard quantum mechanics as Maudlin-type CAEs present to TI.

2. The Maudlin Contingent Absorber Experiment

Let us first review Maudlin's CAE, illustrated in Figure 1.

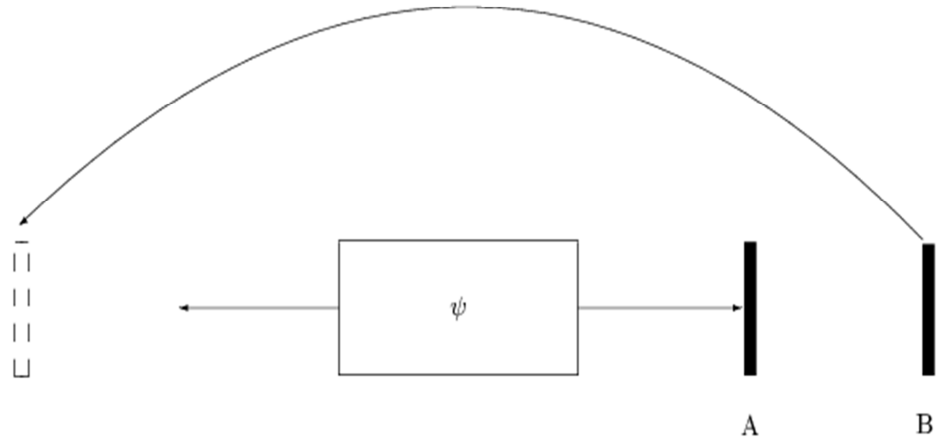

Figure 1. The Maudlin experiment

As illustrated in Figure 1, a source emits massive (and therefore, Maudlin assumes, slow-moving) particles either to the left or right, in the state $|\Psi\rangle = \frac{1}{\sqrt{2}}[|R\rangle + |L\rangle]$, a superposition of 'rightward' and 'leftward' -propagating states. OW components corresponding to right and left are emitted in both directions, but, in Maudlin's setup, only detector A can initially return a CW (since B is blocked by A). If the particle is not detected at A (meaning that the rightward transaction failed), a light signal is immediately sent to detector B, causing it to swing quickly around to intercept the particle on the left. B then is able to return a CW of amplitude $\frac{1}{\sqrt{2}}$. At this point, the particle is certain to be detected there, so Maudlin claims that the CW amplitude of less than unity is evidence of inconsistency on the part of TI. He also argues that the 'pseudotime' picture, in which CW are returned from all absorbers and an 'echoing' process takes place, cannot account for this experiment, since the outcome of the first incipient transaction (i.e., the overlap of the OW and CW component between the emitter and the nearby absorber) must be decided without a CW from the more distant absorber. This appears to undercut the original 1986 account, which assumes that all CW are received by the emitter at once and the choice of which transaction is realized is made based on a sort of "competition" between clearly defined incipient transactions.[1]

Maudlin's challenge has two main features: (1) it seems to involve a situation not amenable to the usual 'echoing' picture as given in Cramer (1986) -- in particular, there seems to be no definite account of what CW are present at the start of the experiment; and (2) it is not clear that the probabilities are consistent or well-defined. Regarding (1), the problem, according to the usual way of thinking, seems to be the following. At t=0 the OW is emitted, but since CW propagate from absorbers back to the emitter, whatever CW are generated at t=2 must 'already' be back at the emitter. Berkovitz (2002) explicitly augments the emitter with a label corresponding to the presence (or absence) of a CW (and the particular state of the CW), calling this the 'state of the emitter'.[2] Depending on the outcome, there are two different 'states of the emitter,' so there seems to be no 'fact of the matter' about which one is the 'correct' one—in contrast to the standard case, discussed in Cramer (1986), in which all absorbers are present to the initial OW. Regarding (2), Maudlin argues that, once A has failed to detect the particle, it is certain to be detected at B, implying that the B outcome should have a probability of 1; but the OW/CW incipient transaction corresponding to detection at B has a weight of only ½ (this corresponds to the Born Rule of standard quantum mechanics and is simply obtained in TI as the product of the amplitudes of the OW and CW components). He therefore argues that TI's probabilities, given by the weights of the associated incipient transactions, seem inconsistent.

The short response to objection (2) is that the weights of incipient transactions are *physical* in nature rather than epistemic (i.e., they are not based on knowledge or ignorance of an observer). In Maudlin's experiment, detection at B will only occur in ½ the trials, and that is what the weight of ½ describes; not the observer's knowledge, based on failure of detection at A in any particular trial, that the particle will be detected at B. We return to (2) in a little more detail below.

Objection (1) is the issue more likely to seen as problematic for TI: it seems to thwart the idea of a clearly defined competing set of incipient transactions as described in Cramer (1986), since one or more absorbers are not available to the emitter unless another outcome occurs (or fails to occur). This seems to set up causal loops, in Berkovitz' terminology, as follows: At t=0 the OW is emitted; at t=1 a CW is returned from *A*, so the 'state of the emitter' can be represented as OW(A). Now suppose the transaction between the emitter and *A* fails. Then a signal is sent from *A* to swing *B* into position to intercept the OW on the left at t=2. Absorber *B* now returns a CW, so the 'state of the emitter' at t=0 is OW(A,B). But here's the causal loop: if the state of the emitter is OW(A,B), it is already certain at t=0 that *B* is in place and the particle must be detected on the left. On the other hand, if the state of the emitter is OW(A), then it is certain at t=0 that the particle must be detected on the right. We seem to have two causal loops with nothing to which we can apply the probabilities of ½ ; moreover, the loops seem to contradict each other: both seem 'predestined' as of t=0, but they obviously can't both happen. [3]

---

[1] Elitzur (2009, private communication) proposes a more elaborate setup in which a CW from a faraway absorber is contingent on the direction of a nearby mirror which itself is activated (or not) by a nearby outcome. Miller (2011, private communication) proposes a photon being split by a beam splitter into two components , one of which is 'trapped' by a set of mirrors and its ultimate CW made contingent on the success or failure of a transaction with a nearby absorber (the 'trapped' component may have a mirror inserted which diverts it to an alternate absorption site, based on the failure of the nearby transaction). Berkovitz (2008) proposes an EPR-type experiment in which a more distant detector's setting depends on the outcome of a nearby measurement.

[2] TI doesn't adopt this ontology, since an emitter is not described by a particular CW, but I understand Berkovitz to be attempting to be precise about the circumstances surrounding CW generation, so I'll use his terminology in this discussion.

Cramer (2005) has replied to such challenges by proposing that the incipient transaction with the shortest spacetime interval has ontological priority; that is, in 'pseudotime' the shorter incipient transaction has its chance to succeed or fail 'before' the longer incipient transaction has a chance. But this cannot be used for cases like the Miller (2011) experiment involving only photons (since all photon spacetime intervals are zero).[4] In the Miller experiment, a photon is split by a half-silvered mirror into two beams A and B; the beam in B is 'trapped' by a fixed set of mirrors to delay its absorption by absorber *B*. If it is not detected at A at t=1, a moveable mirror is quickly inserted into the beam going to detector *B* such that the OW component in that arm is diverted to detector *B'* (perhaps with different properties such as a polarization filter). This makes the specific CW component corresponding to arm B, and returned to the emitter, dependent on the outcome at t=1. Thus there appears to be 'no fact of the matter' about the 'state of the emitter' – i.e., which CW are present, *B* or *B'*.

Before considering the resolution of these (1)-type issues, let's return to objection (2) to see in more detail how TI's probabilities are indeed well-defined in CAE. In the Maudlin experiment, if there is really no other absorber for the OW component heading toward the left, theoretically there may be no incipient transaction on the left, but we may still define the relevant probabilities by taking the total sample space as consisting of the outcomes (Yes, No) for the question "Is the particle detected on the right"? That is, actualization of the incipient transaction between the emitter and A is the answer "Yes" and its failure is the answer "No." Each answer's probability is ½. This is a natural step of applying the law of the excluded middle to cases in which there is only one incipient transaction corresponding to a given outcome for a particular observable: the only physical possibilities are that the one possible transaction either succeeds or that it does not.

---

[3] But see Kastner (2006) which argues that the probability of ½ applies to each loop, and that the emitter state (which is the one that seems to be self-contradictory) should not be viewed as the 'branch point' between loops, but rather the incipient transaction between the emitter and the fixed absorber which then determines which of the possible 'emitter states' is actualized. This implies that the past need not be viewed as determinate, and is basically equivalent to the solution proposed herein. An alternative solution by Marchildon (2006) argues that CW are well-defined in such experiments in a 'block world' picture, which in our view amounts to a 'hidden variable' approach; i.e., that there is always a fact of the matter about the 'state of the emitter' but it is unknown to experimenters. Since I take quantum mechanics as complete, I view the status of quantum objects as genuinely indeterminate.

[4] Strictly speaking, it could be argued that in a direct action theory (upon which TI was originally based), all photons are 'virtual' and can therefore have finite (if nearly zero) rest mass. I neglect that aspect here, but it could be a route to saving the 'hierarchy' approach.

For the more general case, note that we can do a 'spectral decomposition' of any observable, i.e., we can express it in terms of a sum of the projectors onto its eigenstates. In TI these mathematical projectors represent physical incipient transactions, and the density operator for any emitted OW in terms of the absorbers actually available to it corresponds physically to the set of weighted transactions.[5] The projectors constitute a complete disjoint set covering all possible outcomes, and defining a Boolean (classical) probability space. Suppose there are N projectors corresponding to N possible outcomes. Label them each *n, (n=1,N)* where the weight of the *nth* incipient transaction, corresponding to the probability of its associated event being actualized, is P(n).[6] Then P(1) + P(2) + …. +P(N-1) + P(N) = 1. If there are incipient transactions corresponding to $N-1$ of the outcomes but not for a particular outcome *k*, then the probability that none of the N-1 incipient transactions is actualized is *1- $\Sigma_{n \neq k}$ P(n) =P(k)*. This is the generalization of the two-outcome experiment discussed above, in which the probability of the answer "No" to the question "Is the particle on the right?" is the same as the probability of the answer "Yes" to the question "Is the particle on the left"? Thus it is clear that the probabilities for various outcomes can be unambiguously defined for contingent absorber experiments or simply for experiments in which there is not a complete set of absorbers.

3. Delayed choice as a challenge for orthodox quantum mechanics

However, as promised, I return to the issue of an apparent conflict between the 'states of the emitter' in the different apparent causal loops presented by such experiments. The way to deal with this issue is by noticing that a similar conundrum already appears in standard, orthodox quantum theory in delayed-choice experiments. Here is where I take note of the observation by Stapp (2011) that, even in orthodox QM, there is no 'fact of the matter' about certain aspects of the past relative to the times of such delayed choices.[7] To see this, let us briefly review the DCE.

The classic presentation of the DCE is as follows (see Figure 2): (1). At t=0, a photon is emitted towards a barrier with two slits. (2) At t=1, the photon passes the barrier (i.e., we discard runs in which the photon is blocked by the barrier). (3) The photon continues on to a screen *S* on which one would expect to record (at t=2) an interference pattern as individual photon detections accumulate. (4) However, the screen may be removed before the photon arrives (but after it has passed the slit barrier), revealing two telescopes focused on each slit. (5) If this happens, the two telescopes *T* will perform a 'which-slit' measurement at t=3, and the photon will be detected at one or the other telescope, indicating that it went through the corresponding slit (i.e. there is no interference). The decision as to whether to remove *S* or not is made randomly by the experimenter.

---
[5] Specifically, in the expression |x><x|, '|x>' represents the component of the OW absorbed by a detector corresponding to property *x* of observable *X*, and '<x|' represents the CW response of the absorber. If the OW and CW amplitudes are $a_x$ and $a^*_x$ respectively, the associated weighted incipient transaction is represented by $a^*_x a_x$ |x><x|. Note that the set of weighted transactions corresponds to von Neumann's "Process 1" or 'choice of the observer as to what to measure,' the physical origin of which remains mysterious in standard interpretations but which has an obvious physical interpretation in TI.
[6] The weight is the Born Rule, which in TI is simply the product of the amplitudes of the OW and CW comprising each incipient transaction; see the previous note.
[7] This work does not address the aspect of Stapp (2011) which considers the possibility of alteration of the statistical predictions of orthodox quantum theory. It deals only with Stapp's observations concerning standard quantum theory.

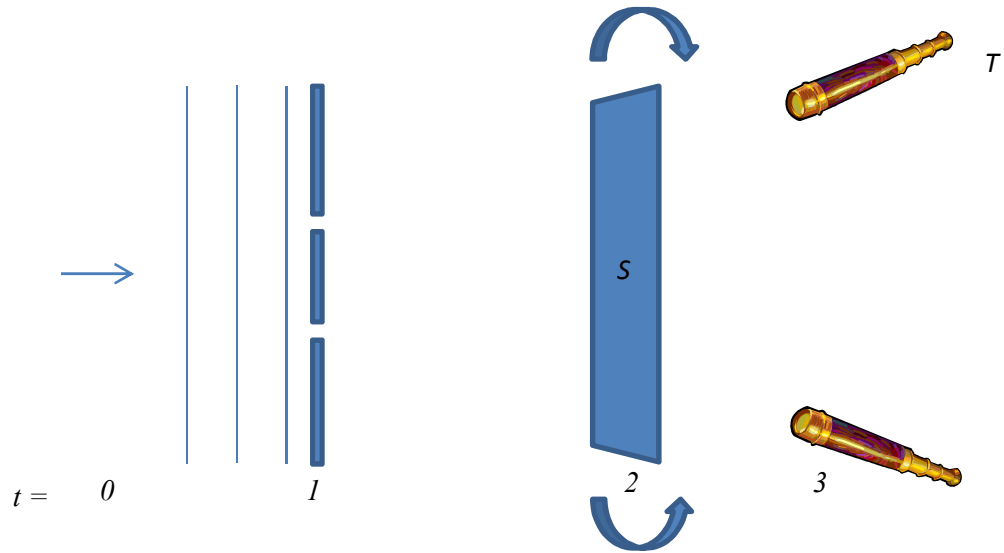

Figure 2: The Delayed Choice Experiment

Note that the photon has already passed the plane of the slits before the observer has decided whether to measure 'which slit' or not. Thus, at a time *1< t <2* prior to the observer's choice, *there is apparently no 'fact of the matter' about the photon's state, including whether or not it has 'interfered with itself.'* The reader might object that the photon's state is simply a superposition of slits, $|\psi\rangle = (1/\sqrt{2})\,[|A\rangle + |B\rangle]$ , but that tacitly assumes that no 'which slit' measurement takes place and implies that interference is present, which is actually uncertain. To see this clearly, we need to take into account situations involving both the preparation *and post-selection* of quantum systems. It is well known that a particle pre-selected in one state and and post-selected in another state (where these can be states of two noncommuting observables) can be equally well described by either the pre- or post-selected state. This is because standard quantum mechanics gives a probability of unity for the outcome corresponding to either the pre- or post-selection state if a measurement of the corresponding observable were conducted at a time between the two states.[8] A photon pre-selected in $|\psi\rangle$ as above, and post-selected via a 'which slit' measurement at t=3 (say yielding the outcome 'slit *A*') can be described by either $|\psi\rangle$ or $|A\rangle$ at t=1.5; but the same photon for which the screen *S* is not removed *cannot be described at any time by a one-slit state*. Thus the photon's ontological status is undefined in a way that goes beyond the usual quantum indeterminacy--i.e., as exemplified by ordinary superpositions such as the state $|\psi\rangle$, or by ambiguous states based on a *specified* pre- and post-selection as discussed above and in footnote 6. In the case of the DCE, we have ambiguity not just based on a given pre- and post-selection but on *an uncertainty in the post-selection itself,* which translates into an essentially different kind of ambiguity in the ontological status of the photon between measurements. [9]

---

[8] This is the standard time-symmetric ABL Rule (1964) for a measurement performed in between a pre- and post-selection. The inference that either the pre- or post-selected state could be attributed to a particle at an intervening time is a direct consequence of the calculus of probabilities applying to standard quantum theory and does not run afoul of any illegitimate counterfactual usage of the ABL Rule. See Kastner (1999), eqs. 24-26 and supporting discussion, for why this is so.

The reader may still think that the above indeterminacy of the photon state is just the usual (relatively benign) quantum indeterminacy. However, this is not the case; the delayed choice experiment also presents a 'causal loop' problem for standard quantum mechanics, as follows. In the usual language, at t=1 a photon progresses past the slit plane. If an experimenter later (at t=2) removes *S* to reveal a 'which slit' detector, this action means that the photon *must have only gone through one or the other slit at t=1*, since there can be no interference between paths corresponding to each slit. [10] According to the 'block world' way of thinking, this means (for logical and physical consistency) that the experimenter *must* place the 'which slit' detector and his choice never was free but was predetermined, even though there was no causal mechanism forcing his choice.

Now, one might just conclude that this implies that there is no genuine freedom of choice and the comparison between the two types of experiments (delayed choice vs, contingent absorber) ends there. However, we can sharpen the 'causal loop' aspect of the delayed choice experiment by proposing that the experimenter does not choose but instead bases removal of *S* on the outcome of a quantum coin flip; say, the outcome of the measurement of 'up' or 'down' of an electron in a Stern-Gerlach (S-G) apparatus. That is, the S-G measurement goes on independently but alongside the usual DCE, and a particular outcome occurring at *t=1.5*, say "up," is used to automatically trigger removal of *S*. Then the outcome of that 'coin flip' *must also be predetermined* based on the effect of the delayed measurement which retroactively decides whether or not the photon 'interfered with itself' while passing through the slits prior to that measurement. Yet quantum mechanics predicts that the outcome of the S-G "coin flip" measurement is uncertain (e.g., has only a 50-50 chance). This is essentially the same alleged inconsistency presented for TI by CAE-type experiments.

---

[9] In the TI picture of the DCE, the photon's OW is perfectly well-defined; it is only its CW that is not well-defined. This is arguably a simpler way to understand the DCE.

[10] Although it might be argued that the Copenhagen interpretation would not countenance *any* statement about the whereabouts of the photon prior to the choice, the usual approach to the DCE, and certainly Wheeler's approach, has been to infer that the choice of a 'which way' measurement determines what happens to the photon in the past. This is the whole point of Wheeler's amplification of the DCE to astronomical proportions as in his version with a photon wavefunction traveling from a distant galaxy and being split by gravitational lensing: to emphasize how present choices may affect an arbitrarily distant past. One might also try to avoid the discussed ontological uncertainty of the photon through instantaneous von Neumann 'collapse', but this approach results in inconsistency regarding the ontological status of the photon for two different inertial observers, for whom such collapses will define different hyperplanes of simultaneity.

To summarize: in the TI case, the alleged inconsistencies presented by the CAE are (1) an apparent lack of 'fact of the matter' concerning which CW are present at t=0 (i.e., 'the state of the emitter' is uncertain) and (2) the apparent causal loop paradox in which one or the other outcome must be predetermined, while the outcome of the incipient transaction in the fixed portion of the experiment is supposed to be uncertain, with a probability of ½ of 'yes' or 'no'. In standard quantum mechanics and the DCE augmented by a quantum coin flip, the inconsistency concerns the apparent lack of 'fact of the matter' about the ontological status of the photon , based on the uncertainty of its post-selection state, for times $1 < t < 2$; and the apparent causal loop paradox in which a particular measurement choice is mandated for the photon, vs. the prediction of 50-50 for the outcome of the quantum coin flip at $t=1.5$ that determines the choice (removal of *S* or not). Note, however, that neither the ontological status of the photon prior to the choice nor the 'state of the emitter' in the TI picture are empirically accessible, so there can be no violation of causality for either case in the form of a "bilking paradox" or other overt contradiction with experience.

Thus we see that the delayed choice experiment (especially when augmented to make the choice dependent on a quantum outcome) seems to raise the same sort of causal loop conundrum that has been used as a basis for criticism of TI. The point here is that *the possibility of an ontologically indeterminate situation at t that becomes determinate in virtue of a later outcome at $t + \Delta t$ is a feature of orthodox quantum mechanics itself* and therefore cannot be viewed as a defect of a particular interpretation such as TI. This should perhaps not be so surprising, since there is no contradiction with observation: the portions of the past that are indeterminate are empirically inaccessible. They become determinate based on events that select out of the possible past events certain actual ones, as discussed by Stapp (2011).

CAEs gain traction as apparent threats to TI based on the idea that there must be a 'fact of the matter' about what CW exist *at the present time of the emission*—that is, when $t=0$ is 'Now'---and such experiments clearly make that impossible . But in fact there is no reason to demand this, since there can likewise be no 'fact of the matter' about the self-interference status of a photon when $t=1.5$ is 'Now' in the DCE in orthodox quantum mechanics.[11] (The two can be legitimately compared because CW in TI are no more empirically accessible than is the ontological status of an individual photon between measurements.) Furthermore, a variant of the DCE raises the same types of consistency issues for standard quantum probabilities (in the usual implicit 'block world' picture) as do CAE for TI. The message of quantum mechanics is that the determinacy of certain aspects of the past depends on what happens in the present (i.e., the future of the past event(s) in question). Once one allows for this metaphysical possibility, the apparent inconsistencies vanish in both cases. The status of a photon's self-interference can be uncertain and contingent on a later measurement, just as the 'state of the emitter' in TI can be uncertain and contingent on the outcome of a later incipient transaction.

---

[11] This locution obviously implies an "A-series" view of time. This aspect will be more fully addressed in a separate work. For purposes of this presentation, it may be noted that the "B-series" or 'block universe' view is part of the problem in that it implies causal loops that need not exist in an "A-series" picture. That is, the photon's status may be indeterminate at t=1.5 (Now=1.5) but determinate at t=1.5 (Now=3), where Now is indexed by Stapp's "process time." The first *t* index can be thought of the number of a row in a knitted fabric, while the second "Now" index can be thought of as indicating which row of stitches is currently on the needle. There need not be any overt conflict between this picture and McTaggart's much-contested 'proof' that time itself does not exist. We need only make a distinction between (i) the numbering of events with respect to each other and (ii) the inception of each event.

4. Conclusion

Contingent absorber experiments (CAE) used as 'counterexamples' to the TI picture present essentially the same conceptual challenge as do delayed choice experiments (DCE) for standard quantum mechanics, so the former therefore cannot legitimately be used as a basis for rejection of TI. Standard quantum mechanics, not just TI, entails that certain features of the past are indeterminate and can only become determinate based on events occurring in the present. This ontological indeterminacy applies to the photon in the delayed choice experiment just as much as it does to TI's confirmation waves.